\documentclass[twocolumn,english]{autart}

\usepackage[T1]{fontenc}
\usepackage[latin9]{inputenc}
\usepackage[a4paper]{geometry}
\usepackage{color}
\usepackage{booktabs}
\usepackage{amsmath}
\usepackage{amssymb}
\usepackage{graphicx}
\usepackage{wasysym}
\usepackage{subscript}

\makeatletter

\providecommand{\tabularnewline}{\\}

\newtheorem{theorem}{Theorem}
\newtheorem{remark}{Remark}
\newtheorem{definition}{Definition}
\newtheorem{algo}{Algorithm}
\newtheorem{method}{Method}

\newcommand{\calX}{{\mathcal{X}}}

\newcommand{\sis}{\setlength{\itemsep}}
\newcommand{\erem}{\hfill $\star$}

\makeatother

\usepackage{babel}
\begin{document}

\begin{frontmatter}{}

\title{Nonlinear system identification in Sobolev spaces}

\author{Carlo Novara, Angelo Nicol\`{\i}, Giuseppe C. Calafiore}

\address{Dept. of Electronics and Telecommunications,
Politecnico di Torino, Italy\\Email: carlo.novara@polito.it, angelo.nicoli@polito.it,
giuseppe.calafiore@polito.it\\Corresponding author: C. Novara}

\begin{abstract}
We consider the problem of deriving from experimental data an approximation
of an unknown function, whose derivatives also approximate the unknown
function derivatives. Solving this problem is useful, for instance,
in the context of nonlinear system identification for obtaining models
that are more accurate and reliable than the traditional ones, based
on plain function approximation. Indeed, models identified by accounting
for the derivatives can provide a better performance in several tasks,
such as multi-step prediction, simulation, Nonlinear Model Predictive
Control, and  control design in general. In this paper, we propose
a novel approach based on convex optimization, allowing us to solve
the aforementioned identification problem. We develop an optimality
analysis, showing that models derived using this approach enjoy suitable
optimality properties in Sobolev spaces. The optimality analysis also
leads to the derivation of tight uncertainty bounds on the unknown
function and its derivatives. We demonstrate the effectiveness of
the approach with two numerical examples. The first one is concerned
with approximation of an univariate function, the second one with
multi-step prediction of the Chua chaotic circuit. 
\end{abstract}

\end{frontmatter}{}

\section{Introduction}

Consider a nonlinear discrete-time system, represented in the following
input-output regression form: 
\begin{eqnarray}
y_{t+1} & = & f_{o}\left(x_{t}\right)+\xi_{t+1}
\label{eq:sys0}
\\ \rule{.8cm}{0cm}
\rule{0cm}{0.4cm}
x_{t} & = & \left(y_{t},\ldots,y_{t-m_{u}+1},u_{t},\ldots,u_{t-m_{u}+1}\right) \nonumber
\label{eq:sys0}
\end{eqnarray}
where $u_{t}\in U\subset\mathbb{R}^{n_{u}}$ is the input, $y_{t}\in\mathbb{R}^{n_{y}}$
is the output, $\xi_{t}\in\varXi\subset\mathbb{R}^{n_{\xi}}$ is a
disturbance and $t\in\mathbb{Z}$ is the discrete time index. The
sets $U$ and $\varXi$ are compact with non-empty interior. The {\em regression
function} $f_{o}$ is supposed to be unknown: the objective of this
paper is to obtain from a batch of experimental data an estimate $\hat{f}$
of $f_{o}$ such that ({\em i}) $\hat{f}$ approximates $f_{o}$,
and ({\em ii}) the first derivatives of $\hat{f}$ approximate
the first derivatives of $f_{o}$. Some relevant motivations for considering
this problem are given next.

\emph{Multi-step prediction and simulation.} A standard approach to
the identification of system (\ref{eq:sys0}) is to adopt a parametrized
NARX (Nonlinear Auto Regressive with eXogenous inputs) model structure
and to estimate the involved parameters by minimizing the model prediction
error; see, e.g., \cite{LLju1,Ljung99}. A relevant issue is that
a model identified using this approach may be accurate when used for
one-step ahead prediction but poor when used for multi-step prediction
or simulation. This may happen, for example, when the model sampling
time is too short; \cite{Goodwin2010}. In this case, the identified
model tends to become a so-called persistent model, where the prediction
is close to the current value: $\hat{y}_{t+1}\cong y_{t}$. Clearly,
a persistent model cannot provide a decent performance when used for
multi-step prediction or simulation. In general, the main reason behind
these kind of issues is that the model just aims to minimize the one-step
prediction error, without really trying to capture the relation between
the output and the individual components of $x_{t}$ and $y_{t+1}$.
An approach that may help overcoming these issues consists in adopting
a NOE (Nonlinear Output Error) model structure, in which the involved
parameters are estimated by minimizing the model simulation error,
see, e.g., \cite{LLju1,Ljung99}. NOE models are often more accurate
than NARX models in multi-step prediction and simulation but require
a higher computational burden, since minimization of the simulation
error is in general a hard nonlinear and non-convex problem. In any
case, also for NOE models there are no guarantees that the relation
between the components of $x_{t}$ and $y_{t+1}$ is correctly captured.
The function derivatives express up to first order precisely these
relations, hence approximating them, together with the system function
$f_{o}$, is crucial in determining an accurate model for control
purposes. 

\emph{Nonlinear Model Predictive Control (NMPC).} NMPC is a widely
used technique for controlling complex nonlinear plants, see, e.g.,
\cite{nmpc_book_1,nmpc_book_2,nmpc_book_3}. Data-driven versions
of this technique can be found in \cite{Piga_8698829,SALVADOR2018356,MANZANO2018505,NoFo16,NOVARA2019417}.
NMPC is based on two main operations: (\emph{i}) multi-step prediction
of the plant behavior, and (\emph{ii}) synthesis of a control law
via on-line optimization, based on the predicted behavior. Clearly,
the availability of an accurate multi-step prediction model is of
paramount importance in NMPC. In particular, at every time $t$, given
the input and output regressors $(u_{t-1},\ldots,u_{t-m_{u}+1})$
and $(y_{t},\ldots,y_{t-m_{u}+1})$, the model should correctly describe
the variations of the predicted output $\hat{y}_{t+\tau}$, $\tau\geq1$,
due to variations of the command input sequence $(u_{t},\ldots,u_{t+\tau-1})$.
As discussed in the previous paragraph, the function derivatives describe
these variations to first order, and this, again, motivates the need
in a control context of approximating the system function $f_{o}$
together with its derivatives. 

\emph{Control sensitivity}. The above considerations are not limited
to NMPC. In general, 
when estimating a regression model  that is to be used for control, e.g., of the type
$
\hat y_{t+1} = \hat f(y_{t},\ldots,y_{t-m_u+1},u_{t},\ldots,u_{t-m_u+1})
$,
it is of paramount importance to capture the sensitivities of the output with respect to the commands $u_{t},\ldots,u_{t-m_u+1}$, and these are given,  to first order approximation, by the derivatives of $\hat f$ w.r.t.\ these variables. Failing to get these sensitivities with sufficient precision may result in a model that responds to commands in a poor way.
%

\begin{remark}\rm 
Although the following one is an elementary fact, it is 
perhaps important to remark that a good uniform error bound on a function's values need not imply a good error bound on the sensitivities (derivatives). Indeed, suppose that we have
$
\hat f(x) = f(x) + e(x)
$,
where $f$ is the true function, $\hat f$ is the identified approximation, and $e$ is the error term. If $\hat f$ is approximated in a standard way, we may have that, over a given domain $\calX$,
$
|e(x)| \leq \epsilon$, $\forall x\in\calX
$,
that is, a uniform bound $\epsilon$ on the absolute approximation error $|\hat f(x) - f(x)|$. The point, however, is that even if $\epsilon$ is small, the error on the sensitivity can be arbitrarily large. We have that 
$
\frac{d \hat f}{dx} = \frac{d f}{dx}  + \frac{d e}{dx}
$,
whence
$
\left| \frac{d \hat f}{dx} - \frac{d f}{dx}  \right| = \left| \frac{d e}{dx}  \right|,
$
and indeed it suffices to consider an example with
$
e(x) = \epsilon \sin (\omega x)
$,
to see that $|e(x) |\leq \epsilon$ for all $x$, but 
$
\left|  \frac{d e}{dx}\right|  =  \epsilon \omega  |\cos (\omega x)|
$,
thus the error on the sensitivity can be arbitrarily large, for arbitrarily large $\omega$.
\erem
\end{remark}

\emph{Related literature.} 
The literature appears to be quite scarce on the topic of approximating
from data a function and its derivatives. The existing methods are
based on different classes of approximators, including radial basis
functions \cite{Mai-Duy2003}, neural networks \cite{XiCa11,Pukrittayakamee2011,Avrutskiy2018},
and deep neural networks \cite{Czarnecki2017}. The numerical results
presented in these papers clearly show that using the information
about the function derivatives leads to significant improvements of
the model accuracy and generalization capabilities. This literature
is interesting and effective in showing the potential of techniques
relying on derivative identification. However, only a limited number
of works carry out a theoretical analysis about the approximation
properties of these techniques \cite{Hornik1990,XiCa11,Czarnecki2017},
and the provided results are often non-constructive, in the sense
that they just prove existence of the required approximating function.
Moreover, the works we found typically assume that the function derivative
samples are available but this may not be true in practical applications.
Also, we observe that the existing techniques allow for the identification
of a model, but they do not provide a description of the uncertainty
associated with this model and its predictions.

\emph{Main Contributions.} 
In this paper, we propose a novel identification approach addressing
all the mentioned issues. The approach allows the identification of
a function together with its derivatives, and it is completely based
on convex optimization. We develop a theoretical optimality analysis,
showing that models obtained using the proposed approach enjoy certain
optimality properties in Sobolev spaces. The optimality analysis also
leads to the derivation of tight uncertainty bounds on the unknown
function and its derivatives, quantifying the modeling error and the
prediction uncertainty. The approach uses samples of the regressor
$x_{t}$, of the function output $y_{t}$ and of the function derivative
outputs. As already mentioned, these latter samples may be not available
in a real-world application. Thus, we further propose a technique
for estimating derivative samples from the function input-output data.
We finally present two numerical examples.
These examples
show that the approach may provide significantly more accurate and
reliable models than the traditional ones based on plain function
approximation (i.e., identified without considering the derivatives).

\emph{Paper orgnization.} 
In Section~\ref{sec:notation},
the notation used in the paper and some basic notions about functional
norms and spaces are introduced. In Section~\ref{sec:prob_form},
the identification problem of interest is formalized. In Section~\ref{sec:Identification-algorithms},
two methods are derived for the joint function and derivatives
identification problem. The optimality properties of these methods
are analyzed in Section~\ref{sec:Optimality-analysis}. Based on
this analysis, tight uncertainty bounds are provided in Section~\ref{sec:Ubounds}.
In Section~\ref{sec:der_values}, an algorithm is proposed for estimating
the derivative values, starting from the function input-output values.
Section~\ref{sec:Numerical-examples} presents the numerical examples.
The conclusions are given in Section~\ref{sec:Conclusions}. All
the theorem proofs can be found in the Appendix.

\section{Notation and preliminaries}

\label{sec:notation}

A column vector $x\in\mathbb{R}^{n_{x}\times1}$ is denoted by $x=\left(x_{1},\ldots,x_{n_{x}}\right)$.
A row vector $x\in\mathbb{R}^{1\times n_{x}}$ is denoted by $x=\left[x_{1},\ldots,x_{n_{x}}\right]=\left(x_{1},\ldots,x_{n_{x}}\right)^{\top}$,
where $\top$ indicates the transpose. The $\ell_{p}$ norm of a vector
$x=\left(x_{1},\ldots,x_{n_{x}}\right)$ is defined as usual. The
2-norm (maximum singular value norm) of a matrix $\Phi\in\mathbb{R}^{m\times n}$
is denoted aby
$\| \Phi\| _{2}$, and 
the $\infty$
is denoted by $\| \Phi\| _{\infty}\doteq \max_{i=1,\ldots,m}\sum_{j=1}^{n}\left|\Phi_{ij}\right|$.
The $\mathcal{L}_{p}$ norm of a function with domain $X\subseteq\mathbb{R}^{n_{x}}$
and codomain in $\mathbb{R}$, is defined as 
$\| f\| _{p}\doteq
\left[\int_{X}\| f\left(x\right)\| _{p}^{p}dx\right]^{\frac{1}{p}}$,
for 
$p\in(1,\infty)$, and 
as 
$\mathrm{ess}\sup_{x\in X}\| f\left(x\right)\| _{\infty}$
for $p=\infty$.
These norms give rise to the well-known $\ell_{p}$ and $\mathcal{L}_{p}\equiv\mathcal{L}_{p}(X)$
Banach spaces.
The $\mathcal{S}_{1p}$ Sobolev norm of a differentiable function
with domain $X\subseteq\mathbb{R}^{n_{x}}$ and codomain in $\mathbb{R}$,
is defined as 
$\| f\| _{\mathcal{S}p}  \doteq\sum_{i=0}^{n_{x}}\| f^{(i)}\| _{p}\label{eq:sob_norm}$,
where 
$
f^{(i)}\doteq 
f$ for $i=0$, and
$
f^{(i)}\doteq 
\frac{\partial f}{\partial x_{i}}$ for $ i>0$.
Note that the superscript $(i)$, with $i>0$, here denotes the partial
derivative of a function with respect to the $i$-th variable, and
not the $i$-th order derivative. The Sobolev norm  gives
rise to the $\mathcal{S}_{1p}\equiv\mathcal{S}_{1p}(X)$ Sobolev space,
also denoted in the literature with $W_{1p}$ or $W_{1,p}$.

\begin{definition}The Sobolev space $\mathcal{S}_{1p}(X)$ is the
set of all functions $f\in\mathcal{L}_{p}(X)$ such that, for every
$i>0$, the derivative $f^{(i)}$ exists and $f^{(i)}\in\mathcal{L}_{p}(X)$:
$
\mathcal{S}_{1p}(X)\doteq\left\{ f:f^{(i)}\in\mathcal{L}_{p}(X),\,i=0,\dots,n_{x}\right\} 
$.
\end{definition} 
Sobolev norms (and related spaces)
involving higher order derivatives can also be found in the literature.
The concept of weak derivative, which is a generalization of the standard
derivative, is often used. In this paper, the interest is for the
case of first order standard derivatives.

\section{Problem formulation}

\label{sec:prob_form}

Consider a function $f_{o}\in\mathcal{S}_{1p}(X)$, taking values
$
z=f_{o}(x)\label{fo}$,
where $x\in X\subset\mathbb{R}^{n_{x}}$, $X$ is a compact set, and
$z\in\mathbb{R}$. Suppose that $f_{o}$ is not known, but a set of
noise-corrupted input-output data from the unknown function is available:
\begin{equation}
D=\left\{ \tilde{x}_{k},\left\{ \tilde{z}_{k}^{i}\right\} {}_{i=0}^{n_{x}}\right\} _{k=1}^{L}\label{eq:id_data}
\end{equation}
where $\tilde{x}_{k}\in X$ are the measurements of the function argument,
$\tilde{z}_{k}^{0}\equiv\tilde{z}_{k}$ are the measurements of the
function output and $\tilde{z}_{k}^{i}$, $i>0$, are the measurements
of the $i$-th partial derivative output. The data (\ref{eq:id_data})
can be described by 
\begin{equation}
\begin{aligned}\tilde{z}_{k}^{i} & =f_{o}^{(i)}\left(\tilde{x}_{k}\right)+d_{k}^{i},\;i=0,\ldots,n_{x},\;k=1,\ldots,L,\end{aligned}
\label{Dset}
\end{equation}
where $d_{k}^{i}$ are noises and $d_{k}^{0}\equiv d_{k}$. If the
data are generated by the system (\ref{eq:sys0}), we have that $\tilde{z}_{k}^{0}\equiv\tilde{z}_{k}=\tilde{y}_{k+1}$,
and the noise terms account for the disturbance $\xi_{t}$ and possible
measurement errors.

We remark that in real-world applications, only the output of the
function is usually measured, while the outputs of the derivatives
may not be available. This more realistic situation is dealt with
in Section~\ref{sec:der_values}, where an algorithm is presented
for estimating the derivative output samples $\tilde{z}_{k}^{i}$,
$i>0$, from the input-output function samples $\tilde{x}_{k}$ and
$\tilde{z}_{k}$.

Now, assume that the noise sequences $d^{i}=(d_{1}^{i},\ldots,d_{L}^{i})$
are unknown but bounded: 
$\| d^{i}\| _{q}\leq\mu^{i}\label{bnois}$,
where $\| \cdot\| _{q}$ is a vector $\ell_{q}$
norm and $0\leq\mu^{i}<\infty$. In the case $q=2$, it can be convenient
to write $\mu^{i}$ as $\mu^{i}=\sqrt{L}\breve{\mu}^{i}$, with $0\leq\breve{\mu}^{i}<\infty$.
In some situations, the noise bounds $\mu^{i}$ are known from the
physical knowledge about the system of interest and the involved sensors.
In other situations, these bounds are not known and have to be estimated
from the available data. An algorithm will be provided in Section
\ref{sec:Ubounds} for performing this estimation.

In this paper, we consider the problem of identifying from the data
(\ref{eq:id_data}) an ``accurate'' approximation $\hat{f}$ of
the unknown function $f_{o}$, such that also the derivatives $\hat{f}^{(i)}$,
$i>0$, of $\hat{f}$ are ``accurate'' approximations of the derivatives
$f_{o}^{(i)}$, $i>0$, of $f_{o}$. The accuracy is measured by means
of the following identification error 
$
e(\hat{f})\doteq \| f_{o}-\hat{f}\| _{\mathcal{S}p}\label{ee}
$,
where $\| \cdot\| _{\mathcal{S}p}$ is a Sobolev
norm. In other words, we are looking for an approximation of the unknown
function $f_{o}$ in the $\mathcal{S}_{1p}$ Sobolev space. Besides
the goal of obtaining such an approximation, we also aim at evaluating
guaranteed estimate bounds for $f_{o}$.

A parametrized structure is adopted for the approximating function:
\begin{equation}
\hat{f}\left(x\right)=\sum_{j=1}^{N}a_{j}\phi_{j}\left(x\right)\label{bfe}
\end{equation}
where $\phi_{j}\in\mathcal{S}_{1p}(X)$ are given basis functions and $a_{j}\in\mathbb{R}$
are coefficients to be identified. The choice of the basis functions
is clearly an important step of the identification process, see, e.g.,
\cite{LLju1,Novara2011711}. In several cases, the basis functions
are known from the physical knowledge of the system of interest. In
other cases the basis functions are known a priori to belong to some
``large'' set of functions, see, e.g., the examples presented in
Section \ref{subsec:Chua} and in \cite{NoACC11}. In yet other cases,
the basis functions are not known a priori and their choice can be
carried out by considering the numerous options available in the literature
(e.g., Gaussian, sigmoidal, wavelet, polynomial, trigonometric, etc.);
see \cite{LLju1} for a discussion on the main features of the most
used basis functions and guidelines for their choice.

The problem considered in this paper is stated as follows.

\begin{prob}\label{id_prob}From the data set $D$ in (\ref{eq:id_data}),
identify an estimate $\hat{f}$ of the form (\ref{bfe}), such
that: 
\begin{itemize}
\item[(i)] the Sobolev identification error $e(\hat f)$ is small; 
\item[(ii)] the estimate is equipped with guaranteed uncertainty bounds on the
unknown function $f_{o}$ and its derivatives.
\end{itemize}
\end{prob}

In the reminder of the paper, for numerical conditioning reasons,
we assume that the components of $x$ in $z=f_{o}(x)$ have similar
ranges of variation. This assumption can always be met through a suitable
rescaling of the components.

\section{Identification methods}

\label{sec:Identification-algorithms}

In this section we propose two  methods for
solving Problem \ref{id_prob}, both based on convex optimization. 
In Section~\ref{sec:Optimality-analysis}
it will be shown that functions identified by means of these methods
enjoy suitable optimality properties. In this section, we suppose
that the derivative output samples $\tilde{z}_{k}^{i}$, $i>0$ are
available. In Section~\ref{sec:der_values}, we will show how these
derivative samples can be estimated from the input-output function
samples $\tilde{x}_{k}$ and $\tilde{z}_{k}$.

A simple yet fundamental observation is that the approximating function
(\ref{bfe}) and its derivatives are given by 
\begin{equation}
\hat{f}^{(i)}\left(x\right)=\sum_{j=1}^{N}a_{j}\phi_{j}^{(i)}\left(x\right),\;i=0,\ldots,n_{x}.\label{eq:bfe-der}
\end{equation}
On the basis of this observation we can present the first identification
method.

\begin{method}\label{algo_1} \textcolor{white}{-} 
\begin{enumerate}
\item Define 
\begin{equation}
\begin{array}{c}
\tilde{z}^{i}\doteq\left[\begin{array}{c}
\tilde{z}_{1}^{i}\\
\vdots\\
\tilde{z}_{L}^{i}
\end{array}\right],\;\Phi^{i}\doteq\left[\begin{array}{ccc}
\phi_{1}^{(i)}(\tilde{x}_{1}) & \cdots & \phi_{N}^{(i)}(\tilde{x}_{1})\\
\vdots & \ddots & \vdots\\
\phi_{1}^{(i)}(\tilde{x}_{L}) & \cdots & \phi_{N}^{(i)}(\tilde{x}_{L})
\end{array}\right]\end{array}.\label{eq:z_phi}
\end{equation}
\item Estimate the vector $a=(a_{1},\ldots,a_{N})$ of model coefficients
in (\ref{eq:bfe-der}) by solving the following convex optimization
problem: 
\begin{align}
 & a=\arg\min_{\alpha\in\mathbb{R}^{N}}\| \alpha\| _{r}\label{eq:alg1}\\
 & \mathrm{s.t.}\;\| \tilde{z}^{i}-\Phi^{i}\alpha\| _{q}\leq\mu^{i},\;i=0,\ldots,n_{x},\label{eq:alg1_con}
\end{align}
where integers $r,q$ indicate suitable vector norms.
\end{enumerate}
\end{method}

The rationale behind this method can be explained as follows: the
constraints (\ref{eq:alg1_con}) ensure that the resulting model (\ref{eq:bfe-der})
is consistent with the available information on the noises corrupting
the data. If the optimization problem is
not feasible, it means that either the chosen basis function set is
not sufficiently rich or the noise bounds $ \| d^{i}\| _{q}\leq\mu^{i}$ are too
small. The minimization of the coefficient vector $\ell_{r}$ norm
in (\ref{eq:alg1}) is carried out for regularization reasons, allowing
also to limit the issue of overfitting. Typical norms that can be
used are the $\ell_{2}$ and $\ell_{1}$ norms. In particular, the
$\ell_{1}$ norm allows one to obtain a sparse coefficient vector
$a$ (see, e.g., \cite{Fuchs05,Tib96,Tropp06,Donoho06_2}), resulting
in a low-complexity model. This is an important property, especially
in view of the model implementation on real-time processors.\smallskip{}

We now present the second identification method.

\begin{method}\label{algo_2} \textcolor{white}{-} 
\begin{enumerate}
\item Define $\tilde{z}^{i}$ and $\Phi^{i}$ as in (\ref{eq:z_phi}). 
\item Estimate the vector $a=(a_{1},\ldots,a_{N})$ of model coefficients
in (\ref{eq:bfe-der}) by solving the following convex optimization
problem: 
\begin{align}
a & =\arg\min_{\alpha\in\mathbb{R}^{N}}\sum_{i=0}^{n_{x}}\lambda^{i}\| \tilde{z}^{i}-\Phi^{i}\alpha\| _{q}^{2}+\Lambda\| \alpha\| _{r}\label{eq:alg2}
\end{align}
where 
where integers $r,q$ indicate suitable vector norms,
and $\lambda^{i}\geq 0,\Lambda\geq 0$ are given weights.
\end{enumerate}
\end{method}

Problem (\ref{eq:alg2}) is aimed at minimizing a tradeoff between the model fitting
error on the identification data and 
a regularization term. For $r=1$, (\ref{eq:alg2})
is a  Lasso problem, see, e.g., \cite{Tib96}; 
for $r=2$, it becomes a classical Ridge regression problem, see, e.g., \cite{Gruber:98}.
Note
that, for suitable values of the parameters $\mu^{i}$, $\lambda^{i}$
and $\Lambda$, the optimization problems (\ref{eq:alg1}) and (\ref{eq:alg2})
are equivalent to each other.

\begin{remark}\rm
It is  worth to stress  the fact that Method~\ref{algo_1}
and Method~\ref{algo_2} are here considered in terms of the guarantees they provide for the ensuing models, and that this paper's contribution lies in the specific models that lead to Sobolev space identification through Method~\ref{algo_1}
and Method~\ref{algo_2}, and in their analysis, and  {\em not} in the actual numerical solution of problems in \eqref{eq:alg1_con} or \eqref{eq:alg2}. These problems indeed have 
a well-known regularized regression structure, and a  pletora of efficient numerical methods already exist for their solution. \erem
\end{remark}

\section{Optimality analysis}

\label{sec:Optimality-analysis}

In Section \ref{sec:Identification-algorithms}, two identification
methods have been presented, allowing us to derive parameterized
approximations of the unknown function $f_{o}$. In this section,
following a Set Membership approach \cite{MiVi91}, \cite{MiNorLaWa96},
\cite{Schweppe73}, \cite{ChGu00}, \cite{Milanese20112141}, \cite{SzWeCaHwTAC09},
we show that such approximations enjoy suitable optimality properties
in Sobolev spaces. Two cases are covered: in the first one, we suppose
that the true function $f_{o}$ belongs to a Sobolev space $\mathcal{S}_{1p}$;
in the second one, we make an additional assumption, regarding the
Lipschitz continuity of the derivatives of the function $f_{o}-\hat{f}$,
which allows us to prove stronger optimality properties of the approximations
with respect to the first case.\textcolor{red}{{} }The analysis and
results developed here are extensions to Sobolev spaces of those regarding
approximation in $\mathcal{L}_{p}$ spaces presented in \cite{MiNoAUT04,Milanese20112141}.

\subsection{Optimality analysis in Sobolev spaces}

\label{subsec:Basic-optimality-analysis}

Consider that the function $f_{o}$ and its derivatives are unknown,
while instead we have the experimental information given by (\ref{eq:id_data})
and (\ref{Dset}), and the prior information given by the inclusion
$f_{o}\in\mathcal{S}_{1p}(X)$ and the noise bounds $\| d^{i}\| _{q}\leq\mu^{i}$.
It follows that $f_{o}\in \mbox{FFS}_{\mathcal{S}}$, where $\mbox{FFS}_{\mathcal{S}}$
is the so-called Feasible Function Set, defined below.

\begin{definition} \label{def:ffs_S} The Feasible Function Set $\mbox{FFS}_{\mathcal{S}}$
is defined as 
$
\mbox{FFS}_{\mathcal{S}}\doteq\{f\in\mathcal{S}_{1p}(X):||\tilde{z}^{i}-f^{(i)}\left(\tilde{x}\right)||_{q}\leq\mu^{i}$,
$i=0,\ldots,n_{x}\}$,
where $f^{(i)}\left(\tilde{x}\right)$ $\doteq$ $(f^{(i)}\left(\tilde{x}_{1}\right),\ldots,f^{(i)}\left(\tilde{x}_{L}\right))$.
\end{definition}

In words, the Feasible Function Set is the set of all functions consistent
with the prior assumptions and with the available data. The Feasible
Function Set thus summarizes all the experimental and a-priori information
that can be used for identification. If at least a function exists
that is consistent with the assumptions and the data (i.e., if $\mbox{FFS}_{\mathcal{S}}\neq\emptyset$),
we say that the assumptions are validated. Otherwise (i.e., if $\mbox{FFS}_{\mathcal{S}}=\emptyset$),
we say that the assumptions are falsified; see \cite{MiNorLaWa96,ChGu00}.

\begin{definition} \label{def:assu_val}The prior assumptions are
considered validated if $\mbox{FFS}_{\mathcal{S}}\neq\emptyset$.\hfill{}$\Square$\end{definition}

The following theorem gives a sufficient conditions for prior assumption
validation.

\begin{theorem}\label{thm:assu_val}$\mbox{FFS}_{\mathcal{S}}\neq\emptyset$
if the optimization problem (\ref{eq:alg1})-(\ref{eq:alg1_con})
is feasible.

\end{theorem}

\textbf{Proof.} See the Appendix.\hfill{}$\Square$

If the optimization problem (\ref{eq:alg1})-(\ref{eq:alg1_con})
is not feasible, it means that either the chosen basis function set
is not sufficiently rich or the noise bounds $\| d^{i}\| _{q}\leq\mu^{i}$ are
too small. In the case where reliable noise bounds are available,
a sufficiently rich basis function set has to be found, considering
the numerous options available in the literature (e.g., Gaussian,
sigmoidal, wavelet, polynomial, trigonometric). If no basis functions
are found for which the optimization problem is feasible, a relaxation
of the noise bounds is needed.

In the reminder of the paper, it is assumed that the prior assumptions
are true and, consequently, $f_{o}\in \mbox{FFS}_{\mathcal{S}}$. Under this
assumption, for a given approximation $\hat{g}$ of $f_{o}$, a tight
bound on the identification error $e(\hat{g})$
is given by the following worst-case error.

\begin{definition} We define the worst-case identification error as
$
\mbox{\mbox{WE}}(\hat{g},\mbox{FFS}_{\mathcal{S}})\doteq\sup_{f\in \mbox{FFS}_{\mathcal{S}}}\| f-\hat{g}\| _{\mathcal{S}p}$,
where $\| \cdot\| _{\mathcal{S}p}$ is the Sobolev
norm.
 \end{definition}

An optimal approximation is defined as a function $f_{op}$ which
minimizes the worst-case approximation error.

\begin{definition} \label{def:opt_app}An approximation $f_{op}$
is $\mbox{FFS}_{\mathcal{S}}$-optimal if 
$
\mbox{\mbox{WE}}(f_{op},\mbox{FFS}_{\mathcal{S}})=\inf_{\hat{g}}\mbox{\mbox{WE}}(\hat{g},\mbox{FFS}_{\mathcal{S}})\doteq\mathcal{R}(\mbox{FFS}_{\mathcal{S}})
$, where 
$\mathcal{R}(\mbox{FFS}_{\mathcal{S}})$ is called the \emph{radius of information}
and is the minimum worst-case error that can be achieved on the basis
of the available prior and experimental information.\hfill{}$\Square$

\end{definition}

In other words, an optimal approximation is the best approximation
that can be found on the basis of the available prior and experimental
information (this information is summarized by the Feasible Function
Set). Finding optimal approximations is in general hard and sub-optimal
solutions can be looked for. In particular, approximations called
almost-optimal are often considered in the literature, see, e.g.,
\cite{Traub88}, \cite{MiNorLaWa96}.

\begin{definition} An approximation $f_{\mbox{ao}}$ is {$\mbox{FFS}_{\mathcal{S}}$-almost-optimal
}if 
$
\mbox{WE}(f_{\mbox{ao}},\mbox{FFS}_{\mathcal{S}})\leq2\inf_{\hat{g}}\mbox{WE}(\hat{g},\mbox{FFS}_{\mathcal{S}})=2\mathcal{R}(\mbox{FFS}_{\mathcal{S}})$.
\end{definition}

The following result gives sufficient conditions under which an approximation
(possibly obtained by the methods of Section \ref{sec:Identification-algorithms})
is almost-optimal.

\begin{theorem}\label{thm:opt_app1}Assume that:\\
i) the optimization problem (\ref{eq:alg1})-(\ref{eq:alg1_con})
is feasible. \\
ii) the approximation $\hat{f}$ given in (\ref{bfe})-(\ref{eq:bfe-der})
has coefficients $a_{j}$ satisfying inequalities (\ref{eq:alg1_con}).
\\
Then, the approximation $\hat{f}$ is \emph{$\mbox{FFS}_{\mathcal{S}}$-}almost-optimal.

\end{theorem}

\textbf{Proof.} See the Appendix.\hfill{}$\Square$

This theorem shows that an approximation obtained by Method \ref{algo_1}
is always almost-optimal. Instead, an approximation obtained by Method
\ref{algo_2} is almost-optimal if its coefficients satisfy inequalities
(\ref{eq:alg1_con}).

\subsection{Optimality analysis with Lipschitz information}

As discussed in Section \ref{subsec:Basic-optimality-analysis}, the
function $f_{o}$ and its derivatives are unknown, while instead we
have available the experimental information given by (\ref{eq:id_data})
and (\ref{Dset}), and the prior information given by the inclusion
$f_{o}\in\mathcal{S}_{1p}(X)$ and the noise bounds.
In this section, we make an additional assumption on the Lipschitz
continuity of the derivatives of the so-called residue function $f_{o}-\hat{f}$.
This allows us to prove stronger optimality properties with respect
to those discussed in Section~\ref{subsec:Basic-optimality-analysis}.

The residue function is defined as 
$
\Delta(x)\doteq f_{o}(x)-\hat{f}(x).\label{eq:res_def}
$.
We assume that this function and its derivatives are Lipschitz continuous.
That is, for given Lipschitz constants $\gamma^{i}<\infty$, $i=0,\dots,n_{x}$,
$
\Delta^{(i)}\in\mathcal{L}(\gamma^{i},X)\label{eq:res_fun}
$,
where 
$\mathcal{L}(\eta,X)\doteq  \{f\in\mathcal{S}_{1p}(X):
  \left|f(x)-f(w)\right|\leq\eta\| x-w\| _{\infty}$, $\forall x,w\in X\}$.
This assumption is reasonable, since we already know that $\Delta\in\mathcal{S}_{1p}(X)$,
which implies that $\Delta$ is Lipschitz continuous and its derivatives
are continuous (a slightly weaker condition with respect to Lipschitz
continuity). The constants $\gamma^{i}$ can be estimated from the
available data by means of the procedure presented at the end of this
section.

Under the Lipschitz condition, we have that $f_{o}\in \mbox{FFS}_{\mathcal{L}}$,
where $\mbox{FFS}_{\mathcal{L}}$ is the following Feasible Function Set.

\begin{definition} \label{def:ffs_L} We let 
$
\mbox{FFS}_{\mathcal{L}}\doteq \{f\in\mathcal{S}_{1p}(X):f^{(i)}-\hat{f}^{(i)}\in\mathcal{L}(\gamma^{i},X)$, 
 $ ||\tilde{z}^{i}-f^{(i)}\left(\tilde{x}\right)||_{q}\leq\mu^{i}$, $i=0,\ldots,n_{x}\}$, 
where $f^{(i)}\left(\tilde{x}\right)$ $\doteq$ $(f^{(i)}\left(\tilde{x}_{1}\right),\ldots,f^{(i)}\left(\tilde{x}_{L}\right))$.
\end{definition}

$\mbox{FFS}_{\mathcal{L}}$ is the set of all functions consistent with the
prior assumptions and the available data. Recalling Definition~\ref{def:assu_val},
a result is now presented, giving sufficient conditions for assumption
validation.

\begin{theorem}\label{thm:assu_val2}$\mbox{FFS}_{\mathcal{L}}\neq\emptyset$
if the optimization problem (\ref{eq:alg1})-(\ref{eq:alg1_con})
is feasible.

\end{theorem}

\textbf{Proof.} See the Appendix.\hfill{}$\Square$

To see how the assumption about the Lipschitz continuity of the function
derivatives helps to obtain stronger optimality properties, consider
Definitions \ref{def:ffs_S} and \ref{def:ffs_L}. These definitions
imply that $\mbox{FFS}_{\mathcal{L}}\subseteq \mbox{FFS}_{S}$ and, consequently,
$
\mathcal{R}(\mbox{FFS}_{\mathcal{L}})\leq\mathcal{R}(\mbox{FFS}_{\mathcal{S}})$.
This inequality shows that the Lipschitz continuity assumption yields
a reduction of the worts-case identification error.

The following result gives sufficient conditions, under which an approximation
is almost-optimal, when the Feasible System Set is $\mbox{FFS}_{\mathcal{L}}$.

\begin{theorem}\label{thm:opt_app2}Let the assumptions of Theorem
\ref{thm:opt_app1} hold and the functions $\Delta^{(i)},\,i=1,\dots,n_{x}$,
be Lipschitz continuous. Then, the approximation $\hat{f}$ is \emph{$\mbox{FFS}_{\mathcal{L}}$-}almost-optimal.

\end{theorem}

\textbf{Proof.} See the Appendix.\hfill{}$\Square$

This section is concluded with a procedure for estimating the constants
$\gamma^{i}$ from the available data. The procedure is the following:
\begin{enumerate}
\item \sis{3mm}Let $\Delta\tilde{z}_{k}^{i}\doteq\tilde{z}_{k}^{i}-\hat{f}^{(i)}(\tilde{x}_{k})$.
The values $\tilde{z}_{k}^{i}$, $k=1,\ldots,L$, are either known/measured
or estimated from the data $\left\{ \tilde{x}_{k},\tilde{z}_{k}\right\} _{k=1}^{L}$,
using Algorithm \ref{algo:grad_est} presented next in Section~\ref{sec:der_values}.
\item Let $\Delta\tilde{z}_{k}^{ij}$ be the samples of the $j$th derivative
of $\Delta^{(i)}$. The values $\Delta\tilde{z}_{k}^{ij}$, $k=1,\ldots,L$,
$i,j=1,\ldots,n_{x}$, are estimated from the data $\left\{ \tilde{x}_{k},\tilde{z}_{k}^{i}-\hat{f}^{(i)}(\tilde{x}_{k})\right\} _{k=1}^{L}$,
using Algorithm \ref{algo:grad_est}. Note that the estimation
of the $\Delta\tilde{z}_{k}^{ij}$'s requires the function $f_{o}$
to be locally twice differentiable at the points $\tilde{x}_{k}$,
$k=1,\ldots,L$. 
\item Estimate the Lipschitz constants $\gamma^{i}$, $i=0,\dots,n_{x}$,
as 
\begin{equation}
\begin{aligned}\gamma^{0} & =\nu\max_{k=1,\ldots,L}\| (\Delta\tilde{z}_{k}^{1},\ldots,\Delta\tilde{z}_{k}^{n_{x}})\| _{\infty}\\
\gamma^{i} & =\nu\max_{k=1,\ldots,L}\| (\Delta\tilde{z}_{k}^{i1},\ldots,\Delta\tilde{z}_{k}^{in_{x}})\| _{\infty}
\end{aligned}
\label{eq:gai_est}
\end{equation}
where $\nu\apprge1$ is a coefficient introduced to guarantee a desired
safety level. 
\end{enumerate}
This procedure is based on the observation that the Lipschitz constant
of a differentiable function is an upper bound of the function's gradient
norm, which gives the motivation for (\ref{eq:gai_est}).

\section{Uncertainty bounds}

\label{sec:Ubounds}

In this section, we derive tight uncertainty bounds for the unknown
function $f_{o}$ and its derivatives $f_{o}^{(i)}$, $i=1,\dots,n_{x}$.
These bounds allow us to quantify the modeling error and the prediction
uncertainty. They can be useful in real-world applications for several
purposes, such as robust control design \cite{Koko96}, \cite{Qu98},
prediction interval evaluation \cite{MiNoTAC05}, and fault detection
\cite{Novara_16_a}. Based on the uncertainty bounds, we present
an algorithm allowing us to estimate the noise bounds $\mu^{i}$. The result presented here regarding the uncertainty
bound derivation is an extension of the one in \cite{MiNoAUT04,Milanese20112141,Novara_16_a}
to the case where the bounds are derived not only for a function but
also for its first-order derivatives.

Under the Lipschitz assumption $\Delta^{(i)}\in\mathcal{L}(\gamma^{i},X)$, we can define the following
functions: 
\begin{equation}
\begin{aligned}\overline{\Delta}^{i}(x) & \doteq\min_{k=1,\ldots,L}\left(\overline{h}_{k}^{i}+\gamma^{i}\| x-\tilde{x}_{k}\| _{\infty}\right)\\
\underline{\Delta}^{i}(x) & \doteq\max_{k=1,\ldots,L}\left(\underline{h}_{k}^{i}-\gamma^{i}\| x-\tilde{x}_{k}\| _{\infty}\right)
\end{aligned}
\label{eq:Deltas_bar}
\end{equation}
where $\overline{h}_{k}^{i}=\tilde{z}_{k}^{i}-\hat{f}^{(i)}(\tilde{x}_{k})+\mu^{i}$,
$\underline{h}_{k}^{i}=\tilde{z}_{k}^{i}-\hat{f}^{(i)}(\tilde{x}_{k})-\mu^{i}$
and $i=0,\ldots,n_{x}$.

A result is now presented, providing tight uncertainty bounds in closed
form for the unknown function $f_{o}$ and its derivatives $f_{o}^{(i)}$.
The result holds in the case where the noise is bounded in $\ell_{\infty}$
norm. The case where the noise
is bounded in $\ell_{2}$ norm is discussed afterwards.

\begin{theorem}\label{thm:opt_bou}Let the assumptions of Theorem
\ref{thm:opt_app1} hold and $q=\infty$ in the noise bounds $\| d^{i}\| _{q}\leq\mu^{i}$. Then,
$
\underline{f}^{i}(x)\leq f_{o}^{(i)}(x)\leq\overline{f}^{i}(x)
$,
where 
\begin{equation}
\begin{aligned}\overline{f}^{i}(x) & =\hat{f}^{(i)}(x)+\min\left(\bar{\gamma},\overline{\Delta}^{i}(x)\right)\\
\underline{f}^{i}(x) & =\hat{f}^{(i)}(x)+\max\left(-\bar{\gamma},+\underline{\Delta}^{i}(x)\right)
\end{aligned}
\label{eq:opt_bou}
\end{equation}
and $\bar{\gamma}\doteq\infty$ if $i=0$ or $\bar{\gamma}\doteq\gamma^{0}$
otherwise. Moreover, 
\begin{equation}
\begin{aligned}\overline{f}^{i}(x) & \doteq\sup_{f\in\mathcal{F}^{i}}f(x)\\
\underline{f}^{i}(x) & \doteq\inf_{f\in\mathcal{F}^{i}}f(x)
\end{aligned}
\label{eq:tightb}
\end{equation}
where 
$
\mathcal{F}^{i}\doteq  \{f:f-\hat{f}^{(i)}\in\mathcal{L}(\gamma^{i},X),||\tilde{z}^{i}-f\left(\tilde{x}\right)||_{\infty}\leq\mu^{i}\}$.
\end{theorem}

\textbf{Proof.} See the Appendix.\hfill{}$\Square$

This theorem shows that, for a given $i\in\{0,\ldots,n_{x}\}$, $\overline{f}^{i}$
and $\underline{f}^{i}$ are the tightest upper and lower bounds of
$f_{o}^{(i)}$ that can be defined on the basis of the information
available about $f_{o}^{(i)}$, summarized by the function set $\mathcal{F}^{i}$.
This result is important since it shows that the bounds $\overline{f}^{i}$
and $\underline{f}^{i}$ are tight. Examples of these bounds are reported
in Figures \ref{fig:bounds_uni} and \ref{fig:chua_pred_bounds} below.
Note that improved bounds on $f_{o}^{(i)}$ could be formally defined
under the assumption $f\in \mbox{FFS}_{\mathcal{L}}$ instead of $f\in\mathcal{F}^{i}$.
However, the evaluation of such bounds would be hard from a computational
point of view. On the contrary, the bounds (\ref{eq:opt_bou}) are
written in closed form and are simple to evaluate.

\begin{remark}\rm It can be proven that the function $f_{c}$ defined
as 
$
f_{c}(x)\doteq\frac{1}{2}\left(\overline{f}^{0}(x)+\underline{f}^{0}(x)\right)
$
is an optimal approximation of $f_{o}$ in any $\mathcal{L}_{p}$
space \cite{MiNoAUT04}. However, $f_{c}$ is not an optimal approximation
of $f_{o}$ in a Sobolev space. Indeed, the derivatives of $f_{c}$
are discontinuous and thus are not appropriate for approximating the
derivatives of $f_{o}$, which instead are continuous.\erem
\end{remark}

In the case where the noise is bounded in $\ell_{2}$ norm (i.e.,
$q=2$ in the noise bounds $\| d^{i}\| _{q}\leq\mu^{i}$, Theorem \ref{thm:opt_bou} cannot be applied
as is, since the $\ell_{2}$ norm bound on the sequence gives no information
on how the individual elements $d_{k}^{i}$ are bounded. In order
to overcome this issue, some additional assumption has to be made
on the element-wise boundedness of the noise sequence $d^{i}$. Suppose
that the estimates $\hat{f}^{(i)}$ obtained from some of the two
identification methods in Section \ref{sec:Identification-algorithms}
are sufficiently accurate approximations of the functions $f_{o}^{(i)}$:
$\hat{f}^{(i)}\left(\tilde{x}_{k}\right)\cong f_{o}^{(i)}\left(\tilde{x}_{k}\right)$.
It follows that $d_{k}^{i}$ $=$ $\tilde{z}_{k}^{i}-f_{o}^{(i)}\left(\tilde{x}_{k}\right)$
$\cong$ $\tilde{z}_{k}^{i}-\hat{f}^{(i)}\left(\tilde{x}_{k}\right)$
$\doteq$ $\delta_{k}^{i}$. It is then natural to consider the following
relative-plus-absolute error bound: 
\begin{equation}
\left\vert d_{k}^{i}\right\vert \leq\zeta_{k}^{i}\doteq\zeta_{R}^{i}\left\vert \delta_{k}^{i}\right\vert +\zeta_{A}^{i},\quad k=1,\ldots,L\label{bnoise2}
\end{equation}
where the term $\zeta_{R}^{i}\left\vert \delta_{k}^{i}\right\vert $
accounts for the fact that $d_{k}^{i}\cong\delta_{k}^{i}$ and $\zeta_{A}^{i}$
accounts for the fact that $d_{k}^{i}$ and $\delta_{k}^{i}$ are
not exactly equal. The parameters $\zeta_{R}^{i},\zeta_{A}^{i}\geq0$
have to be taken such that $\zeta_{R}^{i}\mu^{i}+\zeta_{A}^{i}\sqrt{L}\leq\mu^{i}$.
Indeed, if this inequality is satisfied, (\ref{bnoise2}) is consistent
with $\| d^{i}\| _{q}\leq\mu^{i}$, since  $\| d^{i}\| _{2}$ $\leq$ $\zeta_{R}^{i}\mu^{i}+\zeta_{A}^{i}\sqrt{L}$
$\leq$ $\mu^{i}$. Following this indication, $\zeta_{R}^{i}$ and
$\zeta_{A}^{i}$ can be chosen by means of the procedure presented
at the end of this section. Assuming the bound (\ref{bnoise2}), Theorem
\ref{thm:opt_bou} holds, where the functions $\overline{\Delta}^{i}$
and $\underline{\Delta}^{i}$ in (\ref{eq:opt_bou}) are defined as
in (\ref{eq:Deltas_bar}), with $\mu^{i}\rightarrow\zeta_{k}^{i}$.

Now, a procedure for estimating the noise bounds $\mu^{i}$ in 
is proposed, based on the optimal function bounds given in Theorem
\ref{thm:opt_bou}. For a given $i$, consider the case where $\hat{f}^{(i)}(x)=0$,
$\forall x\in X$. Suppose that the Lipschitz constant $\gamma^{i}$
of the function $\Delta^{(i)}\doteq f_{o}^{(i)}-\hat{f}^{(i)}=f_{o}^{(i)}$
has been estimated by means of Algorithm \ref{algo:grad_est} in Section
\ref{sec:der_values}. According to Theorem \ref{thm:opt_bou}, for
some suitable $\mu^{i}\geq0$, the functions $\overline{f}^{i}$ and
$\underline{f}^{i}$ in (\ref{eq:opt_bou}) are upper and lower bounds
of the unknown function $f_{o}^{(i)}$. Clearly, it must hold that
$\overline{f}^{i}(x)$ $>$ $\underline{f}^{i}(x)$, $\forall x\in X$.
The following procedure provides an estimate of $\mu^{i}$ such that
this inequality is met on the measured data.
\begin{enumerate}
\item \sis{3mm}Let $\hat{f}^{(i)}(x)=0$, $\forall x\in X$. 
\item Solve the following optimization problem: 
\begin{equation}
\begin{aligned} & \underline{\mu}^{i}=\min_{\mu^{i}\geq0}\mu^{i}\\
 & \mathrm{s.t.}\;\overline{f}^{i}(\tilde{x}_{k})>\underline{f}^{i}(\tilde{x}_{k}),\;k=1,\dots,L.
\end{aligned}
\label{eq:mu_min}
\end{equation}
\item Estimate the noise bound as 
$\hat{\mu}^{i}=\nu\underline{\mu}^{i}\label{eq:mu_crit}$,
where $\nu\apprge1$ is a coefficient introduced to guarantee a desired
safety level. 
\end{enumerate}
The optimization problem (\ref{eq:mu_min}) can be easily solved since
the decision variable $\mu^{i}$ is scalar and the number of constraints
is finite. Notice that this procedure uses data only to estimate the
noise bounds, and no preliminary approximations of the unknown function
are required.

\section{Estimation of the derivative values}

\label{sec:der_values}

In practical situations, only the output of the function that describes
the system of interest is usually measured, while the outputs of its
derivatives are not. In this section, we propose an algorithm for
estimating the derivative output samples $\tilde{z}_{k}^{i}$, $i>0$,
from the input-output function samples $\tilde{x}_{k}$ and $\tilde{z}_{k}$.

Suppose that the  data 
$
D^{0}=\left\{ \tilde{x}_{k},\tilde{z}_{k}\right\} _{k=1}^{L}\label{eq:id_data-1}$ is available.
The algorithm for estimating the derivative output samples $\tilde{z}_{k}^{i}$,
$i>0$, is the following.

\begin{algo}\label{algo:grad_est}For $k=1,\ldots,L$: 
\begin{enumerate}
\item \label{enu:ind_sel}\sis{2mm}Define the set of indexes 
\[
\Upsilon_{\rho k}\doteq\{j\in\{1,\ldots,L\}:\| \tilde{x}_{j}-\tilde{x}_{k}\| _{2}\leq\rho\}
\]
where $\rho>0$ is a user-defined radius. 
\item Define the following quantities: 
\[
\begin{array}{c}
\tilde{z}_{\rho k}\doteq\left[\begin{array}{c}
\tilde{z}_{j_{1}}-\tilde{z}_{k}\\
\vdots\\
\tilde{z}_{j_{M}}-\tilde{z}_{k}
\end{array}\right],\;\Phi_{\rho k}\doteq\left[\begin{array}{c}
\tilde{x}_{j_{1}}^{\top}-\tilde{x}_{k}^{\top}\\
\vdots\\
\tilde{x}_{j_{M}}^{\top}-\tilde{x}_{k}^{\top}
\end{array}\right]\end{array}
\]
where $\{j_{1},\ldots,j_{M}\}=\Upsilon_{\rho k}$ and $M=\mathrm{card}\Upsilon_{\rho k}$. 
\item \label{enu:loc_lin}Compute 
\begin{align}
g_{k} & =\arg\min_{\mathfrak{g}\in\mathbb{R}^{n_{x}}}\frac{1}{M}\| \tilde{z}_{\rho k}-\Phi_{\rho k}\mathfrak{g}\| _{2}^{2}.\label{eq:opt_grad_est}
\end{align}
\item \label{enu:loc_comp}Estimate the derivative output samples as 
$
\tilde{z}_{k}^{i}=g_{ki}$, $k=1,\ldots,L$, $i=1,\ldots,n_{x}$,
where $g_{ki}$ are the components of $g_{k}$. 
\item In the case where the data are affected by a relevant noise and/or
the data set is not sufficiently large, the estimated gradient sequence
$\{g_{k}\}_{k=1}^{L}$ can be smoothed by means of a suitable anti-causal
discrete-time filter.\hfill{}$\Square$ 
\end{enumerate}
\end{algo}

The idea behind the algorithm is to identify a local linear model
at each point $\tilde{x}_{k}$ (steps \ref{enu:ind_sel}-\ref{enu:loc_lin}).
The gradient of $f_{o}$ is then estimated by taking the gradient
of this local model, whose coefficients are indeed the gradient components
(step \ref{enu:loc_comp}). The following result provides a bound
on the gradient estimation error.

\begin{theorem}\label{thm:der_est}Assume that:

(i) The derivatives $f_{o}^{(i)}$, $i=1,\dots,n_{x}$, are Lipschitz
continuous on $X$.

(ii) For any $\rho>0$, a $M_{0}>0$ exists such that $\frac{1}{M}\Phi_{\rho k}^{\top}\Phi_{\rho k}\succ0$,
$\forall M\geq M_{0}$.

Then, for any $\epsilon>0$, some $M_{0}>0$ and $\rho>0$ exist such
that the gradient estimation error is bounded as 
\begin{equation}
\| \nabla f_{o}(\tilde{x}_{k})-g_{k}\| _{q}\leq2\| \Phi_{\rho k}^{\dagger}\| _{q}\mu^{0}+\epsilon,\quad\forall M\geq M_{0}\label{eq:grad_bou}
\end{equation}
where $\Phi_{\rho k}^{\dagger}\doteq(\Phi_{\rho k}^{\top}\Phi_{\rho k})^{-1}\Phi_{\rho k}^{\top}$
is the pseudo-inverse matrix of $\Phi_{\rho k}$ and $q\in\{2,\infty\}$.

\end{theorem}

\textbf{Proof.} See the Appendix.\hfill{}$\Square$

This theorem can be interpreted as follows. Two main conditions are
sufficient for obtaining a bound on the gradient estimation error.
The first one (assumption (i) in the theorem) is Lipschitz continuity
of the derivatives $f_{o}^{(i)}$, $i=1,\dots,n_{x}$. This assumption
is reasonable, since we already know that $f_{o}\in\mathcal{S}_{1p}(X)$,
which implies that $f_{o}^{(i)}$, $i=1,\dots,n_{x}$, are continuous
(a slightly weaker condition with respect to Lipschitz continuity).
The second one (assumption (ii)) is a standard persistence of excitation
condition \cite{Ljung99,Novara2011711}. The next result shows that,
under these two assumptions and some further technical conditions,
the gradient estimate converges to its true value as $\rho\rightarrow0$
and $M\rightarrow\infty$.

\begin{theorem}\label{thm:der_est2}Let the assumptions of Theorem
\ref{thm:der_est} be true. Let also the following limits hold: 
\begin{align}
 & \lim_{\rho\rightarrow0}\lim_{M\rightarrow\infty}\frac{1}{M}D_{k}^{\top}D_{k}=\sigma_{D}^{2}\label{eq:var}\\
 & \lim_{\rho\rightarrow0}\lim_{M\rightarrow\infty}\frac{1}{M}D_{k}^{\top}\Phi_{\rho k}=0\label{eq:unc}
\end{align}
where $D_{k}\doteq(d_{j_{1}}-d_{k},\ldots,d_{j_{M}}-d_{k})$ and $0\leq\sigma_{D}^{2}<\infty$.
Then, 
$
\lim_{\rho\rightarrow0}\lim_{M\rightarrow\infty}\| \nabla f_{o}(\tilde{x}_{k})-g_{k}\| _{q}=0$. 

\end{theorem}

\textbf{Proof.} See the Appendix.\hfill{}$\Square$

This theorem shows that, in order to ensure convergence of the estimate
to the true gradient, the limits (\ref{eq:var}) and (\ref{eq:unc})
must hold (besides the basic assumptions of Theorem \ref{thm:der_est}).
The limit (\ref{eq:var}) means convergence of the sample noise variance.
The limit (\ref{eq:unc}) implies sample uncorrelation between the
noise and the regressor. Both these limits (in their statistical version)
represent standard assumptions in the literature on system identification,
see, e.g., \cite{Ljung99}.

\section{Numerical examples}

\label{sec:Numerical-examples}

Two numerical examples are presented in this section. The first one
is a very simple example concerning a scalar function of a scalar
variable. The second one discusses a more complex application to multi-step
prediction of the Chua chaotic circuit. These examples show that the
proposed approach may provide significantly more accurate and reliable
models than traditional approaches based on plain function approximation.

\subsection{Example: univariate function approximation}

\label{subsec:univariate} The following univariate function is considered
in this example: 
\[
f_{o}(x)=\sin(1.1x)
\]
where $x\in\mathbb{R}$ and $f_{o}:\mathbb{R}\rightarrow\mathbb{R}$.

The function and its derivative were evaluated in $L=100$ linearly
equally spaced points in the domain $X=[-2,3]$. A normally distributed
noise with zero mean and standard deviation 0.05 was added to both
the function values and its derivative values, computed analytically.
Hence, a noise-corrupted identification dataset of the form \eqref{eq:id_data}
was obtained. A validation dataset of length $L=1000$ was also obtained
in the same domain $X$. This set consists of noise-free data, in
order to compare the output of the models that will be identified
with the true function values.

\begin{figure}
\centering \includegraphics[width=0.9\columnwidth]{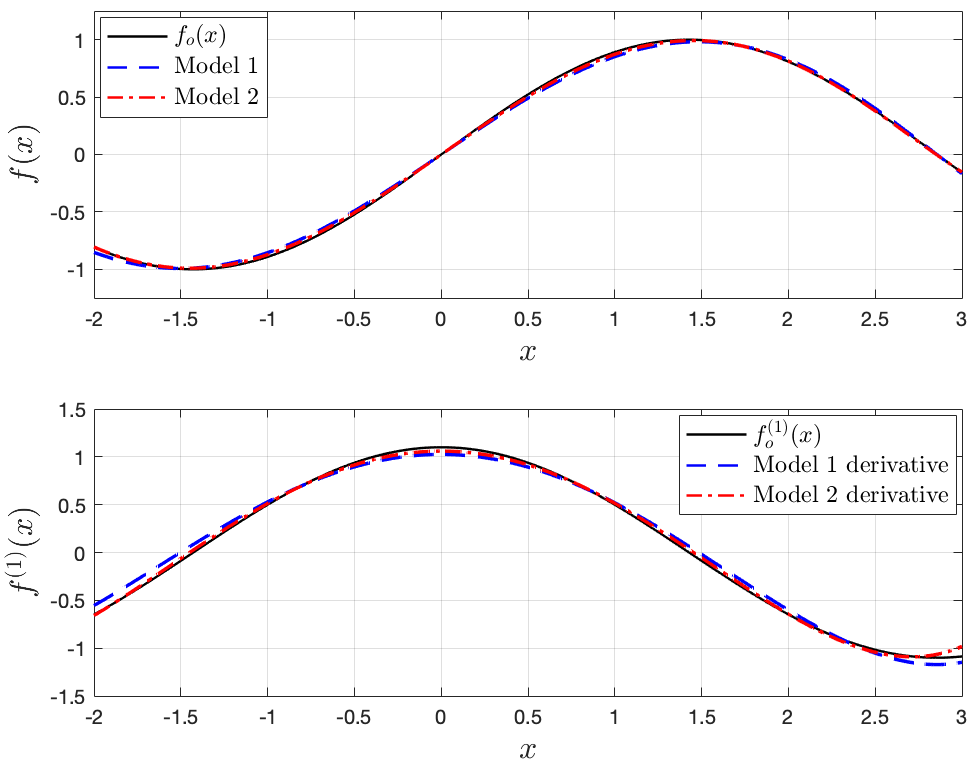}
\caption{Validation set. Upper plot: comparison between true function and model
outputs. Lower plot: comparison between true derivative and model
derivatives. }
\label{fig:f_uni} 
\end{figure}

\begin{figure}
\centering \includegraphics[width=0.9\columnwidth]{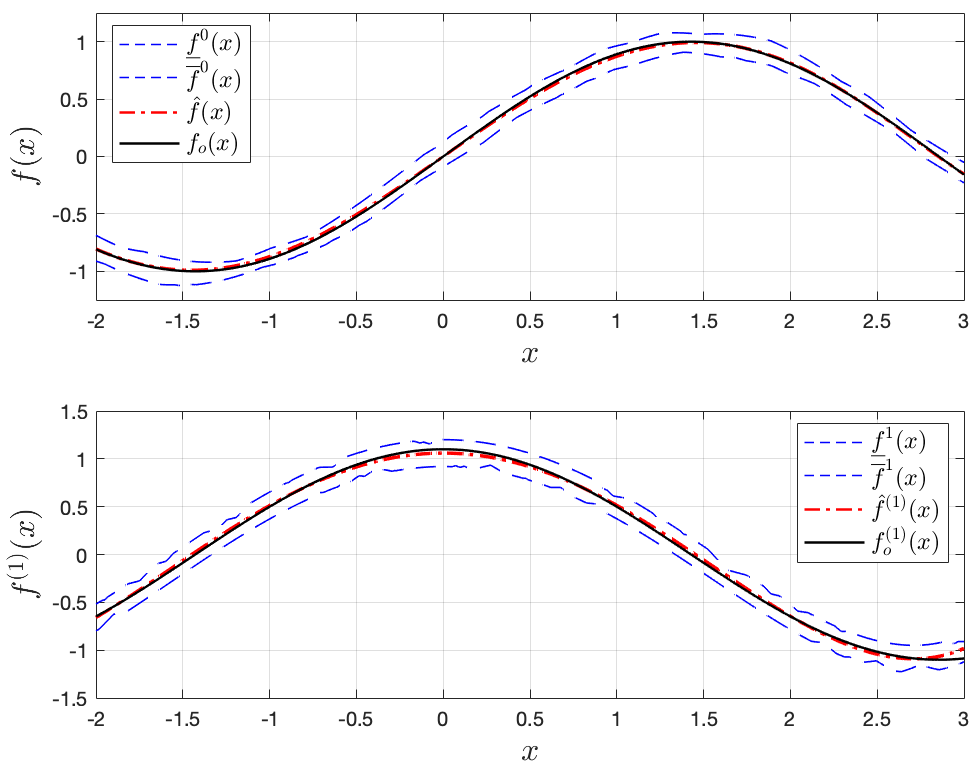}
\caption{Validation set. True function, Model 2 output, derivative and related
uncertainty bounds. }
\label{fig:bounds_uni} 
\end{figure}

A model function of the form \eqref{bfe} was considered, with a basis
function set composed of univariate monomials up to degree $d=5$.
Two models were identified from the identification dataset: 
\begin{itemize}
\item \emph{Model 1}. Function values used for model identification, function
derivative values not used. The coefficients $a_{j}$ in \eqref{bfe}
were identified by Method \ref{algo_2}, with $q=2$, $r=1$, $\lambda^{0}=1$,
$\lambda^{1}=0$, and $\Lambda=1$. \medskip{}
 
\item \emph{Model 2}. Both function and derivative values used for model
identification. The derivative values were computed analytically.
The coefficients $a_{j}$ in \eqref{bfe} were identified by Method
\ref{algo_2}, with $q=2$, $r=1$, $\lambda^{0}=1$, $\lambda^{1}=2$,
and $\Lambda=1$. 
\end{itemize}
The results obtained by the two models on the validation dataset are
summarized in Table \ref{table:RMSE_uni}, where the obtained Root
Mean Square Errors are reported. RMSE is the error between the true
function $f_{o}$ and the model $\hat{f}$; RMSE$^{(1)}$ is the error
between the true function derivative $f_{o}^{(1)}$ and the model
derivative $\hat{f}^{(1)}$.\textcolor{red}{{} }The upper plot in
Figure \ref{fig:f_uni} shows the comparison between the true function
values and the outputs of the identified models. The lower plot in
Figure \ref{fig:f_uni} shows the comparison between the true derivative
values and the outputs of the model derivatives. In Figure \ref{fig:bounds_uni},
the Model 2 uncertainty bounds, computed according to \eqref{eq:opt_bou},
are reported.

From these results, we can conclude that the model identified using
the derivative values (Model 2) provides a more accurate approximation
of the true function derivative with respect to the model identified
not using the derivative values (Model 1). What is quite surprising
is that Model 2 provides also a better approximation of the true function
itself.

\begin{table}[t]
\centering%
\begin{tabular}{lcc}
\toprule 
Estimations  & RMSE  & RMSE$^{(1)}$\tabularnewline
\midrule 
Model 1  & 2.54e-02  & 5.86e-02 \tabularnewline
Model 2  & 1.19e-02  & 2.57e-02 \tabularnewline
\bottomrule
\end{tabular}\caption{RMSE errors on the validation set.}
\label{table:RMSE_uni} 
\end{table}

\subsection{Example: multi-step prediction for the Chua chaotic circuit}

\label{subsec:Chua} The Chua circuit is a simple electronic circuit
showing a chaotic behavior, see \cite{ChKoMa86}. It is composed
of two capacitors, an inductor, a locally active resistor and a nonlinear
resistor. The circuit continuous-time state equations are the following:
\begin{equation}
\begin{aligned}\dot{x}_{1} & =\alpha(x_{2}-x_{1}-\rho(x_{1}))\\
\dot{x}_{2} & =x_{1}-x_{2}+x_{3}+u+\xi^{c}\\
\dot{x}_{3} & =-\beta x_{2}-Rx_{3}\\
y & =x_{1}
\end{aligned}
\label{eq:chua_sys}
\end{equation}
where the states $x_{1}\in\mathbb{R}$ and $x_{2}\in\mathbb{R}$ represent
the voltages across the capacitors, $x_{3}\in\mathbb{R}$ the current
through the inductor, $u\in\mathbb{R}$ is an external input, $y\in\mathbb{R}$
is the system output, $\xi^{c}\in\mathbb{R}$ is a disturbance, and
$\alpha\in\mathbb{R},\,\beta\in\mathbb{R}$ and $R\in\mathbb{R}$
are parameters. In this example, the following nonlinear resistor
characteristic and parameter values are assumed: $\rho(x_{1})=-1.16x_{1}+0.041x_{1}^{3}$,
$R=0.1,\,\alpha=10.4,\,\beta=16.5$. With this parameter values and
nonlinearity, the system exhibits a chaotic behavior and thus prediction
is an extremely hard task.

The system \eqref{eq:chua_sys}, discretized via the forward Euler
method, can be written in the following input-output regression form:
\begin{equation}
\begin{aligned}y_{t} & =b_{1}y_{t-1}+b_{2}y_{t-2}+b_{3}y_{t-3}\\
 & +b_{4}\rho(y_{t-1})+b_{5}\rho(y_{t-2})+b_{6}\rho(y_{t-3})\\
 & +b_{7}u_{t-2}+b_{8}u_{t-3}+\xi_{t}
\end{aligned}
\label{eq:io_form}
\end{equation}
where $\xi_{t}$ is a noise accounting for the disturbance $\xi^{c}$
in \eqref{eq:sys0} and $b_{i}$ are suitable parameters. Equivalently,
it can be written in the form \eqref{eq:sys0}, with $x_{t}=\left(y_{t},y_{t-1},y_{t-2},u_{t-1},u_{t-2}\right)$.

The system \eqref{eq:chua_sys} has been implemented in Simulink.
The input $u$ was simulated as a normally distributed random signal
with zero mean and standard deviation (std) $1$. The disturbance
$\xi^{c}$ was simulated as a normally distributed random signal with
zero mean. Two std values were considered for this disturbance: $0.01$
and $0.05$. For each of these std values, two simulations of duration
$60$ s were carried out and, correspondingly, two set of data of
the form \eqref{eq:id_data} were collected with a sampling time $T_{s}=0.01$
s, corresponding to an experiment length $L=6000$ for every dataset.
The first dataset was used for model identification, the second one
for model validation.

For each std value of the disturbance $\xi^{c}$, the following prediction
models were identified from the identification dataset. 
\begin{itemize}
\item \emph{One-step predictor identified not using any derivative information
(P1\_NOD)}. The predictor P1\_NOD is given by 
\begin{equation}
\begin{aligned}y_{t+1} & =\hat{f}\left(x_{t}\right)\\
x_{t} & =\left(y_{t},y_{t-1},y_{t-2},u_{t-1},u_{t-2}\right)
\end{aligned}
\label{eq:mod_struc}
\end{equation}
where $\hat{f}$ is of the form \eqref{bfe}. A basis function set
composed of multivariate monomials has been used, defined as 
\begin{equation}
\{\phi_{j}\}_{j=1}^{N}=\{\prod_{i=1}^{n_{x}}x_{i,t}^{\alpha_{i}-1};\alpha_{i}=1,2;i=1,\ldots,n_{x}\}\label{eq:mon_set}
\end{equation}
where $x_{i,t}$ is the $i$th component of $x_{t}$ and $n_{x}=5$.
This set consists of $N=2^{n_{x}}=32$ basis functions. The coefficients
$a_{j}$ in \eqref{bfe} were identified by Method \ref{algo_2},
with $q=2$, $r=1$, $\lambda^{0}=1$, $\lambda^{i}=0,\,i>0$, and
$\Lambda=50$. \medskip{}
 
\item \emph{One-step predictor identified using the true derivative values
(P1\_D)}. The predictor P1\_D is of the form \eqref{eq:mod_struc}.
The basis functions are the same as those used in \eqref{eq:mod_struc}.
The true derivative values computed from \eqref{eq:io_form} were
used to construct the vector $\tilde{z}^{i}$, $i>0$, in \eqref{eq:z_phi}.
The coefficients $a_{j}$ in \eqref{bfe} were identified by Method
\ref{algo_2}, with $q=2$, $r=1$, $\lambda^{0}=1$, $\lambda^{i}=200,\,i>0$,
and $\Lambda=50$.\medskip{}
 
\item \emph{One-step predictor identified using the estimated derivative
values (P1\_ED)}. The predictor P1\_ED is of the form \eqref{eq:mod_struc}.
The basis functions are the same as those used in \eqref{eq:mod_struc}.
The derivative values estimated by Algorithm \ref{algo:grad_est}
were used to construct the vector $\tilde{z}^{i}$, $i>0$, in \eqref{eq:z_phi}.
The coefficients $a_{j}$ in \eqref{bfe} were identified by Method
\ref{algo_2}, with $q=2$, $r=1$, $\lambda^{0}=1$, $\lambda^{i}=200,\,i>0$,
and $\Lambda=50$.\medskip{}
 
\item \emph{Direct multi-step predictor identified not using any derivative
information (PK\_NOD)}. The predictor PK\_NOD is given by \vspace{-3mm}
\begin{equation}
\begin{aligned}y_{t+k} & =\hat{f}\left(x_{t}\right)\\
x_{t} & =\left(y_{t},y_{t-1},y_{t-2},u_{t+k-2},u_{t+k-3},\dots,u_{t-2}\right)
\end{aligned}
\label{eq:mod_struc_K}
\end{equation}
where $\hat{f}$ is of the form \eqref{bfe} and $k\in\{3,5,7\}$.
The basis function set is defined as in \eqref{eq:mon_set}, with
$n_{x}=4+k$. This set consists of\textcolor{red}{{} }$N=2^{4+k}$
basis functions. The coefficients $a_{j}$ in \eqref{bfe} were identified
by Method \ref{algo_2}, with $q=2$, $r=1$, $\lambda^{0}=1$,
$\lambda^{i}=0,\,i>0$, and $\Lambda=50$.\medskip{}
 
\item \emph{Direct multi-step predictor identified using the estimated derivative
values (PK\_ED)}. The predictor PK\_ED is of the form \eqref{eq:mod_struc_K}.
The basis functions are the same as those used in \eqref{eq:mod_struc_K}.
The derivative values estimated by Algorithm \ref{algo:grad_est}
were used to construct the vector $\tilde{z}^{i}$, $i>0$, in \eqref{eq:z_phi}.
The coefficients $a_{j}$ in \eqref{bfe} were identified by Method
\ref{algo_2}, with $q=2$, $r=1$, $\lambda^{0}=1$, $\lambda^{i}=200,\,i>0$,
and $\Lambda=50$. 
\end{itemize}
For each std value of the disturbance $\xi^{c}$ (std $\in\{0.01,0.05\}$),
the identified models were tested on the validation set in the task
of $k$-step ahead prediction, with $k\in\{3,5,7\}$. The $k$-step
prediction of models P1\_NOD, P1\_D and P1\_ED was computed by iterating
$k$ times equation \eqref{eq:mod_struc}. The $k$-step prediction
of models PK\_NOD and PK\_ED was computed directly using equation
\eqref{eq:mod_struc_K}.

\begin{figure}
\centering \includegraphics[width=1\columnwidth]{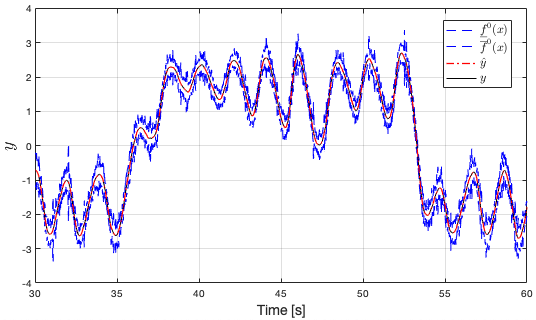}
\caption{Validation set (a portion); std $=0.05$. 3-step prediction of model
PK\_ED and related uncertainty bounds. }
\label{fig:chua_pred_bounds} 
\end{figure}

The results of these tests are summarized in Tables \ref{tab:RMSE_std1}
and \ref{tab:RMSE_std5}, where the Root Mean Square prediction Errors
RMSE\textsubscript{$k$} are reported, for $k\in\{3,5,7\}$ and std
$\in\{0.01,0.05\}$. Figure \ref{fig:chua_pred_bounds} shows the
true system output, the 3-step prediction of the model PK\_ED (in
the case where std $=0.05$) and the related uncertainty bounds, for
a portion of the validation set. Note that these results were obtained
using Method \ref{algo_2}. Similar results can be obtained using
Method \ref{algo_1} (they are not reported here for the sake of
brevity).

The main observation arising from these results is that the models
identified by the proposed method, using the information about
the derivatives, are significantly more accurate (about one order
of magnitude) than those identified not using this information. A
second observation is that the models identified using the estimated
derivative values show a performance similar to those identified using
the true derivative values. A third observation (important in general
but less important than the other two in the context considered in
this paper) is that the direct $k$-step predictors are in general
more accurate than the iterated $1$-step predictors.\smallskip{}

\begin{table}[t]
\centering \label{tab:RMSE_std1} %
\begin{tabular}{lccc}
\toprule 
Predictors  & RMSE\textsubscript{3}  & RMSE\textsubscript{5}  & RMSE\textsubscript{7}\tabularnewline
\midrule 
P1\_NOD  & 6.21e-02  & 1.03e-01  & 1.45e-01\tabularnewline
P1\_D  & 5.55e-03  & 1.27e-02  & 2.29e-02\tabularnewline
P1\_ED  & 3.60e-03  & 1.01e-02  & 2.09e-02\tabularnewline
PK\_NOD  & 6.01e-02  & 9.97e-02  & 1.41e-01\tabularnewline
PK\_ED  & 5.69e-04  & 8.73e-04  & 1.87e-03\tabularnewline
\bottomrule
\end{tabular}\caption{Validation set; std $=0.01$; $k\in\{3,5,7\}$. RMSE prediction errors.}
\end{table}

\begin{table}[t]
\centering \label{tab:RMSE_std5} %
\begin{tabular}{lccc}
\toprule 
Predictors  & RMSE\textsubscript{3}  & RMSE\textsubscript{5}  & RMSE\textsubscript{7}\tabularnewline
\midrule 
P1\_NOD  & 6.18e-02  & 1.03e-01  & 1.44e-01\tabularnewline
P1\_D  & 5.59e-03  & 1.29e-02  & 2.33e-02\tabularnewline
P1\_ED  & 3.56e-03  & 1.01e-02  & 2.09e-02\tabularnewline
PK\_NOD  & 5.96e-02  & 9.90e-02  & 1.39e-01\tabularnewline
PK\_ED  & 6.22e-04  & 1.07e-03  & 2.17e-03\tabularnewline
\bottomrule
\end{tabular}\caption{Validation set; std $=0.05$; $k\in\{3,5,7\}$. RMSE prediction errors.}
\end{table}

\section{Conclusions}

\label{sec:Conclusions}

An approach for the identification of a function together with its
derivatives has been proposed in this paper. Within this approach,
an optimality analysis has been developed, guaranteed uncertainty
bounds have been derived and a technique for estimating the derivative
values from the input-output data has been presented. The approach
has been tested in two simulated examples. One of these examples is
concerned with multi-step prediction of the Chua chaotic circuit.
In this example, the models identified using our methods resulted
to be significantly more accurate than other models obtained using
a standard identification technique, demonstrating the potential of
the proposed identification approach. Future research activities will
regard the derivation of prediction models suitable for the data-driven
NMPC techniques of \cite{NoFo16,NOVARA2019417} and the application
to problems of practical interest.

\section*{Appendix: Theorem proofs}

\textbf{Proof of  Theorems \ref{thm:assu_val} and \ref{thm:opt_app1}.}
If the optimization problem (\ref{eq:alg1})-(\ref{eq:alg1_con})
is feasible, then an approximation $\hat{f}$ of the form (\ref{bfe})
exists, such that inequalities (\ref{eq:alg1_con}) are satisfied.
These inequalities are equivalent to the following ones: $||\tilde{z}^{i}-\hat{f}^{(i)}\left(\tilde{x}\right)||_{q}\leq\mu^{i}$,
$i=0,\ldots,n_{x}$. Moreover, $\hat{f}\in\mathcal{S}_{1p}(X)$ by
definition. It follows that $\hat{f}\in \mbox{FFS}_{\mathcal{S}}$, which
implies $\mbox{FFS}_{\mathcal{S}}\neq\emptyset$. This proves Theorem \textbf{\ref{thm:assu_val}.}

As shown in \cite{Traub88}, \cite{MiNorLaWa96}, if $\hat{f}\in \mbox{FFS}_{\mathcal{S}}$,
then $\hat{f}$ is \emph{$\mbox{FFS}_{\mathcal{S}}$-}almost-optimal. This
proves Theorem \ref{thm:opt_app1}.\hfill{}$\Square$

\textbf{Proof of Theorems \ref{thm:assu_val2} and \ref{thm:opt_app2}.
}The proof of Theorem \ref{thm:assu_val} shows that, if the optimization
problem (\ref{eq:alg1})-(\ref{eq:alg1_con}) is feasible, then an
approximation $\hat{f}$ of the form (\ref{bfe}) exists, and $\hat{f}\in \mbox{FFS}_{\mathcal{S}}$.
Consider now the function $f=\hat{f}+\Delta$, with $\Delta=0$. Obviously,
$f=\hat{f}\in \mbox{FFS}_{\mathcal{S}}$ and $f^{(i)}-\hat{f}^{(i)}=\Delta=0\in\mathcal{L}(\gamma^{i},X)$,
for any $\gamma^{i}\geq0$. From Definitions \ref{def:ffs_S} and
\ref{def:ffs_L}, it follows that $f=\hat{f}\in \mbox{FFS}_{\mathcal{L}}$,
which implies $\mbox{FFS}_{\mathcal{L}}\neq\emptyset$. This proves Theorem
\ref{thm:assu_val2}.

As shown in \cite{Traub88}, \cite{MiNorLaWa96}, if $\hat{f}\in \mbox{FFS}_{\mathcal{L}}$,
then $\hat{f}$ is $\mbox{FFS}_{\mathcal{L}}$-almost-optimal. This proves
Theorem \ref{thm:opt_app2}.\hfill{}$\Square$

\textbf{Proof of Theorem \ref{thm:opt_bou}.} The proof for the case
$i=0$ comes from Theorem 3 in \cite{Novara_16_a}. This theorem
shows that the following bounds hold for every $x\in X$: 
\begin{align*}
f_{o}(x) & \leq\overline{f}^{0}(x)\equiv\hat{f}(x)+\overline{\Delta}^{0}(x)\\
f_{o}(x) & \geq\underline{f}^{0}(x)\equiv\hat{f}(x)+\underline{\Delta}^{0}(x).
\end{align*}
In the case $i>0$, under the assumption (\ref{eq:res_fun}), we can
follow the same argumentations of the proof of Theorem 3 in \cite{Novara_16_a}.
In this way, we obtain that the following bounds hold for every $x\in X$:
\begin{equation}
\begin{aligned}f_{o}^{(i)}(x) & \leq\hat{f}^{(i)}(x)+\overline{\Delta}^{i}(x)\\
f_{o}^{(i)}(x) & \geq\hat{f}^{(i)}(x)+\underline{\Delta}^{i}(x).
\end{aligned}
\label{eq:bound1}
\end{equation}
Moreover, we know that $\Delta^{(0)}$ is Lipschitz continuous with
constant $\gamma^{0}$. This implies that 
\begin{equation}
\left|f_{o}^{(i)}-\hat{f}^{(i)}(x)\right|\leq\gamma^{0}\equiv\bar{\gamma},\;i=1,\ldots,n_{x}.\label{eq:bound2}
\end{equation}
The bounds (\ref{eq:opt_bou}) for $i>0$ are obtained from (\ref{eq:bound1})
and (\ref{eq:bound2}). Equations (\ref{eq:tightb}) follow from Theorem
2 in \cite{MiNoAUT04}.\textcolor{white}{-}\hfill{}$\Square$

\textbf{Proof of Theorem \ref{thm:der_est}.} Let us consider the
Taylor expansion of $f_{o}$ around a point $\tilde{x}_{k}$: 
\begin{align*}
f_{o}(x) & =f_{o}(\tilde{x}_{k})+(x-\tilde{x}_{k})^{\top}\nabla f_{o}(\tilde{x}_{k})+R(x-\tilde{x}_{k})
\end{align*}
where $\nabla f_{o}=(f_{o}^{(1)},\ldots,f_{o}^{(n_{x})})$ is the
gradient of $f_{o}$ and $R(\cdot)$ is a reminder. This expression,
evaluated at a point $\tilde{x}_{j}$, with $j\in\Upsilon_{\rho k}$,
becomes 
\begin{align*}
f_{o}(\tilde{x}_{j}) & =f_{o}(\tilde{x}_{k})+(\tilde{x}_{j}-\tilde{x}_{k})^{\top}\nabla f_{o}(\tilde{x}_{k})+R(\tilde{x}_{j}-\tilde{x}_{k}).
\end{align*}
From (\ref{Dset}), this can be written as 
\begin{align*}
\tilde{z}_{j}-\tilde{z}_{k} & =(\tilde{x}_{j}-\tilde{x}_{k})^{\top}\nabla f_{o}(\tilde{x}_{k})+R(\tilde{x}_{j}-\tilde{x}_{k})+d_{j}-d_{k}.
\end{align*}
For $j=j_{1},\ldots,j_{M}$, we obtain the following equation in matrix
form: 
\[
\tilde{z}_{\rho k}=\Phi_{\rho k}\nabla f_{o}(\tilde{x}_{k})+\Xi_{k}+D_{k}
\]
where $\Xi_{k}\doteq(R(\tilde{x}_{j_{1}}-\tilde{x}_{k}),\ldots,R(\tilde{x}_{j_{M}}-\tilde{x}_{k}))$
and $D_{k}\doteq(d_{j_{1}}-d_{k},\ldots,d_{j_{M}}-d_{k})$. It follows
that 
\[
\nabla f_{o}(\tilde{x}_{k})=\Phi_{\rho k}^{\dagger}\tilde{z}_{\rho k}-\Phi_{\rho k}^{\dagger}(\Xi_{k}+D_{k})
\]
where $\Phi_{\rho k}^{\dagger}\doteq(\Phi_{\rho k}^{\top}\Phi_{\rho k})^{-1}\Phi_{\rho k}^{\top}$.
The inverse $(\Phi_{\rho k}^{\top}\Phi_{\rho k})^{-1}$ exists and
is finite since $\frac{1}{M}\Phi_{\rho k}^{\top}\Phi_{\rho k}\succ0$,
$\forall M\geq M_{0}$, by assumption. This matrix inequality also
implies that the solution of the optimization problem (\ref{eq:opt_grad_est})
is given by 
$
g_{k}=\Phi_{\rho k}^{\dagger}\tilde{z}_{\rho k}$.
The vector $g_{k}$ is an estimate of the gradient $\nabla f_{o}(\tilde{x}_{k})$.
The resulting estimation error $\nabla f_{o}(\tilde{x}_{k})-g_{k}$
is bounded as 
\[
\begin{aligned} & \| \nabla f_{o}(\tilde{x}_{k})-g_{k}\| _{q}=\| \Phi_{\rho k}^{\dagger}(\Xi_{k}+D_{k})\| _{q}\\
 & \leq\| \Phi_{\rho k}^{\dagger}\| _{q}\| \Xi_{k}+D_{k}\| _{q}\leq\| \Phi_{\rho k}^{\dagger}\| _{q}\left(\| \Xi_{k}\| _{q}+2\mu^{0}\right).
\end{aligned}
\]
Being $f_{o}^{(i)}$ Lipschitz continuous by assumption, each element
of $\Xi_{k}$ is bounded as 
\[
\left|R(\tilde{x}_{j}-\tilde{x}_{k})\right|\leq\gamma_{R}\| \tilde{x}_{j}-\tilde{x}_{k}\| _{q}\leq\rho\gamma_{R},\,\forall\tilde{x}_{j}\in X
\]
for some $\gamma_{R}\geq0$, $\gamma_{R}<\infty$. It follows that,
for $\forall M\geq M_{0}$, 
\begin{equation}
\| \Xi_{k}\| _{q}\leq\left\{ \begin{array}{ll}
\rho\sqrt{M}\gamma_{R}, & q=2\\
\rho\gamma_{R}, & q=\infty.
\end{array}\right.\label{eq:Deltak_bou}
\end{equation}
Hence,
\[
\begin{aligned} & \| \nabla f_{o}(\tilde{x}_{k})-g_{k}\| _{q}\leq\| \Phi_{\rho k}^{\dagger}\| _{q}\| \Xi_{k}\| _{q}+\| \Phi_{\rho k}^{\dagger}\| _{q}+2\mu^{0}\\
 & \leq\| \Phi_{\rho k}^{\dagger}\| _{q}\rho\sqrt{M}\gamma_{R}+\| \Phi_{\rho k}^{\dagger}\| _{q}+2\mu^{0}\;(q=2)\\
\mathrm{or} & \leq\| \Phi_{\rho k}^{\dagger}\| _{q}\rho\gamma_{R}+\| \Phi_{\rho k}^{\dagger}\| _{q}+2\mu^{0}\;(q=\infty).
\end{aligned}
\]
The statement is proven choosing $\rho=\epsilon/(\| \Phi_{\rho k}^{\dagger}\| _{q}\sqrt{M}\gamma_{R})$
($q=2$) or $\rho=\epsilon/(\| \Phi_{\rho k}^{\dagger}\| _{q}\gamma_{R})$
($q=\infty$).\hfill{}$\Square$

\textbf{Proof of Theorem \ref{thm:der_est2}.} Let us denote the function
gradient as $g_{o}\doteq\nabla f_{o}(\tilde{x}_{k})$ and, for a certain
gradient estimate $g$, the estimation error as $\delta g\doteq g_{o}-g$.
The objective function of the optimization problem (\ref{eq:opt_grad_est})
is 
\[
J(g)\doteq\frac{1}{M}\| \tilde{z}_{\rho k}-\Phi_{\rho k}g\| _{2}^{2}.
\]
This function can be written as 
\begin{align*}
J(g) & =\frac{1}{M}(\tilde{z}_{\rho k}-\Phi_{\rho k}g)^{\top}(\tilde{z}_{\rho k}-\Phi_{\rho k}g)\\
 & =\frac{1}{M}\left(\tilde{z}_{\rho k}-\Phi_{\rho k}g_{o}+\Phi_{\rho k}\delta g\right)^{\top}\left(\tilde{z}_{\rho k}-\Phi_{\rho k}g_{o}+\Phi_{\rho k}\delta g\right)\\
 & =\frac{1}{M}\left(\Xi_{k}+D_{k}+\Phi_{\rho k}\delta g\right)^{\top}\left(\Xi_{k}+D_{k}+\Phi_{\rho k}\delta g\right)\\
 & =\frac{1}{M}\Xi_{k}^{\top}\Xi_{k}+\frac{1}{M}D_{k}^{\top}D_{k}+\frac{1}{M}\delta g^{\top}\Phi_{\rho k}^{\top}\Phi_{\rho k}\delta g\\
 & +\frac{2}{M}D_{k}^{\top}\Xi_{k}+\frac{2}{M}\Xi_{k}^{\top}\Phi_{\rho k}\delta g+\frac{2}{M}D_{k}^{\top}\Phi_{\rho k}\delta g.
\end{align*}
From (\ref{eq:Deltak_bou}) and the noise bounds $\| d^{i}\| _{q}\leq\mu^{i}$, a sufficiently large
$M_{0}$ exists such that 
\[
\begin{array}{l}
\frac{1}{M}\Xi_{k}^{\top}\Xi_{k}\leq\gamma_{R}^{2}\rho^{2},\;\forall M\geq M_{0}\\
\frac{1}{M}\left|D_{k}^{\top}\Xi_{k}\right|\leq2\breve{\mu}^{i}{}^{0}\gamma_{R}\rho,\;\forall M\geq M_{0}.
\end{array}
\]
From (\ref{eq:var}) and (\ref{eq:unc}), for every $\epsilon>0$,
a sufficiently large $M_{0}$ exists such that 
\[
\textnormal{\ensuremath{\begin{array}{l}
\left|\frac{1}{M}D_{k}^{\top}D_{k}-\sigma_{D}^{2}\right|\leq\epsilon,\;\forall M\geq M_{0}\\
\left|\frac{1}{M}D_{k}^{\top}\Phi_{\rho k}\delta g\right|\leq\| \delta g\| _{2}\epsilon,\;\forall M\geq M_{0}.
\end{array}}}
\]
Moreover, 
\[
\frac{1}{M}\left|\Xi_{k}^{\top}\Phi_{\rho k}\delta g\right|\leq\frac{1}{\sqrt{M}}\| \Phi_{\rho k}\| _{2}\| \delta g\| _{2}\gamma_{R}\rho.
\]
The quantity $\| \Phi_{\rho k}\| _{2}/\sqrt{M}$
is bounded as 
\[
\begin{array}{l}
\frac{1}{\sqrt{M}}\| \Phi_{\rho k}\| _{2}\leq\frac{1}{\sqrt{M}}\left(\sum_{j=1}^{M}\sum_{i=1}^{n_{x}}(\Phi_{\rho k})_{ji}^{2}\right)^{1/2}\\
\leq\frac{1}{\sqrt{M}}\left(n_{x}M\max_{i,j}(\Phi_{\rho k})_{ji}^{2}\right)^{1/2}=\sqrt{n_{x}}\max_{i,j}\left|(\Phi_{\rho k})_{ji}\right|
\end{array}
\]
where the first inequality is a standard result in the literature
and $(\Phi_{\rho k})_{ji}$ are the entries of $\Phi_{\rho k}$. Note
that $\max_{i,j}\left|(\Phi_{\rho k})_{ji}\right|$ is bounded, since
the measurements $\tilde{x}_{j}$ are assumed to be in a compact set.
The quantity $\| \delta g\| _{2}$ is bounded on
any compact set $G$ containing $g_{o}$: for all $g\in G$, $\| \delta g\| _{2}\leq\bar{G}$,
for some $\bar{G}>0$, $\bar{G}<\infty$.

From all the above inequalities, we have that 
\[
\begin{aligned}\left|J(g)-J_{o}(g)\right| & \leq\gamma_{R}^{2}\rho^{2}+4\breve{\mu}^{i}{}^{0}\gamma_{R}\rho+\epsilon+2\bar{G}\epsilon\\
 & +2\sqrt{n_{x}}\max_{i,j}\left|(\Phi_{\rho k})_{ji}\right|\bar{G}\gamma_{R}\rho
\end{aligned}
\]
where 
\[
J_{o}(g)\doteq\frac{1}{M}\delta g^{\top}\Phi_{\rho k}^{\top}\Phi_{\rho k}\delta g+\sigma_{D}^{2}.
\]
It follows that, as $\rho\rightarrow0$ and $M\rightarrow\infty$,
$J(g)$ converges to $J_{o}(g)$.\textcolor{red}{{} }

This convergence is uniform on any compact set $G$ containing $g_{o}$.
It follows that the minimizers of $J(g)$ converge to the minimizers
of $J_{o}(g)$, see \cite{Ljung99}. The condition $\frac{1}{M}\Phi_{\rho k}^{\top}\Phi_{\rho k}\succ0$
ensures that $J_{o}(g)$ has a unique minimizer, given by $g_{o}\doteq\nabla f_{o}(\tilde{x}_{k})$.
The claim follows.\hfill{}$\Square$

\bibliographystyle{plain}
\bibliography{lettaltr,lettnos_journals,lettnos_conferences,sparsification,mpc,sobolev,lettnos_others}

\end{document}